\documentstyle[11pt,seceq]{article}

\newcommand{\be}{\begin{equation}}
\newcommand{\ee}{\end{equation}}
\newcommand{\lb}{\label}

\newcommand{\oz}{{\overline{z}}}
\newcommand{\bv}{{\bf v}}
\newcommand{\br}{{\bf r}}
\newcommand{\bk}{{\bf k}}

\newcommand{\hK}{\hat{K}}
\newcommand{\hA}{\hat{A}}
\newcommand{\hp}{\hat{p}}
\newcommand{\hrho}{\hat{\rho}}
\newcommand{\hvepsilon}{\hat{\varepsilon}}
\newcommand{\cL}{{\cal L}}
\newcommand{\cN}{{\cal N}}

\newcommand{\grad}{{\mbox{\boldmath $\nabla$}}}
\newcommand{\bdot}{{\mbox{\boldmath $\cdot$}}}

\begin{document}
\title{Fluctuations in the Irreversible Decay of Turbulent Energy}
\author{Gregory L. Eyink\\{\em Department of Mathematics}\\
{\em University of Arizona}\\{\em Tucson, AZ 85721}}
\date{ }
\maketitle
\begin{abstract}
A fluctuation law of the energy in freely-decaying, homogeneous and isotropic
turbulence
is derived within standard closure hypotheses for 3D incompressible flow. In
particular,
a fluctuation-dissipation relation is derived which relates the strength of a
stochastic
backscatter term in the energy decay equation to the mean of the energy
dissipation rate.
The theory is based on the so-called ``effective action'' of the energy history
and illustrates
a Rayleigh-Ritz method recently developed to evaluate the effective action
approximately within
probability density-function (PDF) closures. These effective actions generalize
the Onsager-Machlup
action of nonequilibrium statistical mechanics to turbulent flow. They yield
detailed, concrete
predictions for fluctuations, such as multi-time correlation functions of
arbitrary order,
which cannot be obtained by direct PDF methods. They also characterize the mean
histories by a
variational principle.
\end{abstract}

\newpage

\section{Introduction}

\noindent We consider here the problem of fluctuations of the energy in high
Reynolds number turbulence decay. The mean
energy decay in homogeneous and isotropic turbulence has been the subject of
many classic investigations. A rather
thorough review is contained in \cite{MY}, Section 16. von K\'{a}rm\'{a}n and
Howarth \cite{vKH} first derived a
power decay law for the mean energy, $K_*(t)\propto (t-t_0)^{-n}$, by means of
a hypothesis of complete
self-preservation of the spectrum. The fundamental paper of Kolmogorov
\cite{K41} rederived that result, with
a precise prediction for the exponent, $n={{10}\over{7}}$. Kolomogorov's
original argument assumed, however, the
conservation of the Loitsyansky invariant, which was later called into question
by Proudman and Reid \cite{PR}.
Nevertheless, Kolmogorov's basic argument may still be carried through under a
weaker hypothesis, a ``principle
of permanence of the large eddies''. This now-standard theory has been
discussed in several books and reviews:
\cite{Orszag}, \cite{Rey}, and \cite{Frisch}. According to this theory, the
decay exponent $n$ is dependent
on the initial-data, through the power of the low-wavenumber part of the
spectrum.

Our interest here is in the fluctuations of the energy history during the
decay, including joint multi-time statistics.
The main results have been briefly announced elsewhere \cite{EyAl}. Our
analysis is based on a general approach to
fluctuations in irreversible processes, proposed by Onsager \cite{Ons31} and
developed in detail by Onsager and Machlup
\cite{OM}. In this method, an ``action functional'' is employed which measures
directly the probability of observing
a given history as a fluctuation event. In particular, the most probable
history minimizes this action functional. In
systems close to thermal equilibrium, there is a standard
fluctuation-dissipation relation for molecular noise, so that
the Onsager-Machlup action has there the physical interpretation of a
``dissipation function''. Onsager's variational
principle reduces near equilibrium to a ``principle of least dissipation'',
generalizing the well-known hydrodynamic
principle of Rayleigh.

In its original form, however, Onsager's principle was restricted to weakly
noisy systems and could not be
applied to turbulence, where fluctuations are large. Recently we have proposed
a generalization which applies
as well to strongly noisy systems \cite{Ey96I,Ey96II}. The variational
functionals in this theory, or ``effective actions'',
have experimental consequences for turbulence fluctuations and are subject to
realizability conditions which arise from
positivity of the underlying statistical distributions. For each random
variable $Z(t)$ in the flow (where $Z$ may represent
a velocity at a chosen point, a pressure, a turbulent energy, etc.) there is a
corresponding effective action
$\Gamma[z]$, which is a functional of the whole time-history $\{z(t): t_0\leq
t< +\infty\}$ of the variable.
The realizability conditions on this action function are {\em (i)} that it be
nonnegative , $\Gamma[z]\geq 0$,
{\em (ii)} that it have the ensemble mean $\oz(t)$ as its unique minimum
$\Gamma[\oz]=0$, and {\em (iii)} that it
be convex, $\lambda\Gamma[z_1]+(1-\lambda)\Gamma[z_2]\geq \Gamma[\lambda
z_1+(1-\lambda)z_2],\,\,0<\lambda<1$.
As a consequence, the mean value $\oz(t)$ is characterized by a {\em principle
of least effective action}.
Like Onsager's action, this functional directly measures the probability of
fluctuations of the sample histories
away from the mean history. The effective action also serves as a generating
functional for (irreducible) multitime
correlation functions of the considered random variable.

To make the effective action into a practical working tool, efficient and
economical approximation procedures are
required to calculate it. In \cite{Ey96I,Ey96II} we have demonstrated one such
scheme, a Rayleigh-Ritz variational
method inspired by the similar ones already extensively used in quantum theory.
This variational method is designed to be
used in conjunction with probability density-function (PDF) closures, such as
mapping closures \cite{CCK,GK}, generalized
Langevin models \cite{HP,Pope93}, etc. Any reasonable guess of the turbulence
statistics may be input into the variational
method to yield approximations of the effective actions. By this means,
predictions are obtained for multi-time statistics
which are not obtainable by direct PDF methods. The additional information
about fluctuations has been found to be
very useful in evaluating the reliability of PDF closures for practical
modelling purposes \cite{EyAl}.

The contents of this paper are as follows: In Section 2 we very briefly review
the standard theory of the mean energy
decay in high Reynolds number turbulence and, in particular, we cast it in the
form of a PDF-based moment closure.
In Section 3 we evaluate the effective action within the standard theoretical
hypotheses, by means of the Rayleigh-Ritz
algorithm. The realizability of the approximate effective action is verified in
the small-fluctuation regime by means of
a Langevin dynamics for the turbulent energy and a fluctuation-dissipation
relation is derived for the strength
of the stochastic noise term. In Section 4 we discuss some of the testable
consequences of the theory.
In particular, the prediction for the 2-time correlation of the turbulent
energy is given. Also, the direct empirical
significance of the effective action is discussed, in terms of fluctuations in
$N$-sample ensemble averages.

\section{Review of Theory for the Mean Decay}

\noindent We outline here the standard theory of mean energy decay in a
freely-decaying homogeneous and isotropic
turbulence with random initial data at high Reynolds number, following the
reviews in \cite{Orszag,Rey,Frisch}. For
convenience, we assume a model energy spectrum
\be E(k,t)= \left\{ \begin{array}{ll}
                     Ak^m   & k\leq k_L(t) \cr
                     \alpha\varepsilon^{2/3}(t)k^{-5/3} & k_L(t)\leq k \leq
k_d(t) \cr
                     0      & k\geq k_d(t)
                     \end{array} \right.
\lb{spectrum} \ee
which has been adopted in some previous studies \cite{C-BC,Rey}. Such a
spectrum may certainly be taken at time
$t=t_0$ for the initial velocity statistics. We are assuming as well that there
is a permanent form of the spectrum,
according to which the spectral shape is unchanged in time except through its
dependence on the parameters
$\varepsilon(t),k_L(t)$ and $k_d(t)$. Note that the spectrum is not
self-preserving, or self-similar, in the
usual sense discussed in \cite{MY}, which would imply that it have the form
$E(k,t)=\alpha\varepsilon^{2/3}(t)k^{-5/3}
f(k \ell(t))$ for some length-scale $\ell(t)$. In fact, the spectrum contains
two distinct length-scales,
the integral or outer scale $L(t)=k_L^{-1}(t)$ and the dissipation or inner
scale $\eta(t)=k_d^{-1}(t).$
Certain features of the above model are crude caricatures of reality. For
example, the spectrum should not vanish
for $k>k_d(t)$ at any time $t>t_0$, even if it did so initially. However, the
spectrum should always show some rapid
exponential decay in the far dissipation range. It may be easily checked that
such refinements do not change any
of the results below. The important assumption has to do with the
low-wavenumber part of the spectrum. It was found by
Proudman and Reid from the quasinormal closure \cite{PR} that there is a
backscatter term $\sim k^4$ in the energy
transfer $T(k,t)$. Hence, as long as $m<4$ one expects that there is a
permanence of the low-wavenumber spectrum,
in the sense that the power-law $k^m$ and its coefficient $A$ remain unchanged
in time. On the other hand, if $m>4$
initially, then it is expected that the spectrum with $m=4$ will be established
at positive times and, if $m=4$ initially,
then the low-wavenumber spectrum will remain of the same form with a
time-dependent coefficient $A(t)$. Here we always
consider $m<4$, so that the ``permanence of large-eddies'' should hold. For
finiteness of the total energy, $m>-1$ must
also be imposed and, in fact, we usually take $m>0$ so that the spectrum
decreases asymptotically at very low wavenumbers.
For grid-generated turbulence it is has been inferred that $m\approx 2$
\cite{C-BC}.

The mean decay law can be derived very simply from these hypotheses. One
relation $k_L(t)=\left({{\alpha}\over{A}}
\varepsilon^{2/3}(t) \right)^{{{3}\over{3m+5}}}$ is imposed on the spectral
parameters by requiring continuity at
$k=k_L(t)$. An additional constraint is obtained at high Reynolds number by
evaluating the dissipation rate
\be \varepsilon(t)= 2\nu\int_0^\infty \,k^2E(k,t)\,dk \lb{dissint}, \ee
which, for $k_L(t)\ll k_d(t)$, leads to
$k_d(t)=\left({{2}\over{3\alpha\nu}}\right)^{3/4}\varepsilon^{1/4}(t)$. Only
one independent parameter is left, which may be taken to be the integral
$E(t)=\int_0^\infty dk E(k,t)$ representing
mean energy at time $t$. For the above form of the spectrum it is not hard to
show that at high Reynolds number, when
$k_L(t)\ll k_d(t)$, the dissipation is given as
\be \varepsilon(t)= \Lambda_m E^p(t) \lb{diss} \ee
with
$\Lambda_m^{-1}=\alpha^{3/2}\left({{1}\over{m+1}}
+{{3}\over{2}}\right)^{{{3m+5}\over{2m+2}}}A^{{{1}\over{m+1}}}$ and
$p={{3m+5}\over{2m+2}}$. Thus, employing the Navier-Stokes equation via its
energy-balance, one obtains the closed
moment equation
\be \dot{E}(t)= -\Lambda_m E^p(t). \lb{energ-eq} \ee
This generalizes the decay equation (29) in the paper \cite{K41} of Kolmogorov.
Its solution with initial condition
$E(t_0)=K_0$ gives a prediction for the energy decay law, in the form
\be K_*(t) = K_0\left({{t-t_0^*}\over{\Delta t}}\right)^{-n} \lb{law} \ee
with $n={{2m+2}\over{m+3}}$. Here $\Delta t\equiv \left[\Lambda_m (p-1)
K_0^{p-1}\right]^{-1}$ is
a constant with units of time, determined by the initial mean energy $K_0$, and
$t_0^*\equiv t_0-\Delta t$ is a
``virtual time-origin''.

This simple theory may be cast into the form of a PDF closure by assuming as an
{\em Ansatz} at all times $t\geq t_0$ a
Gaussian random velocity field with zero mean and with spectrum $E(k,t)$ given
by Eq.(\ref{spectrum}) above. The assumption
of Gaussian statistics was not used in previous works. It is not necessary here
either, but it makes simpler the analytical
labor in applying the Rayleigh-Ritz algorithm. In fact, the Gaussian {\em
Ansatz} will only be used to evaluate averages of
1-point velocity moments, which it is known have statistics in actual fact
close to Gaussian. See \cite{Batch}, Ch.VIII.
It will be shown later that the use of a non-Gaussian {\em Ansatz} would change
only a constant in the final results. The
model spectrum contains one free parameter, which may be taken to be the energy
per mass $E(t)$ in the {\em Ansatz}, or,
what is the same, its mean value of the quadratic velocity-moment functional
$\hK(\br;\bv)={{1}\over{2}}v^2(\br)$ at some
chosen space-point $\br$. By statistical homogeneity, the mean value is
independent of this choice. The time-dependence of
$E(t)$ is then determined by projecting the Navier-Stokes dynamics onto this
single moment function:
\be \dot{E}(t)= \langle \cL^\dagger\hK(\br)\rangle_{E(t)} \lb{project} \ee
where
\be \cL= -\sum_{i=1}^3\int d^3\br \,{{\delta}\over{\delta
v_i(\br)}}\left[\left(-(\bv(\br)\bdot\grad)v_i(\br)-\nabla_i p(\br)
                                          +\nu\bigtriangleup
v_i(\br)\right)\cdots\right] \lb{Liouv} \ee
is the Liouville operator which generates the evolution of PDF's for the
Navier-Stokes dynamics, $\cL^\dagger$ is the
adjoint operator which generates the evolution of observables, and
$\langle\cdots\rangle_{E(t)}$ denotes average
with respect to the model Gaussian velocity with energy $E(t)$. It is easy to
see that this prescription to determine
the time-dependence leads to
\be \dot{E}(t)= -\varepsilon(t) \lb{decayeq} \ee
which, using Eq.(\ref{diss}), is clearly equivalent to the one above. However,
putting the analysis into this form allows
us to apply the Rayleigh-Ritz method of \cite{Ey96I,Ey96II} to evaluate the
effective actions.

\section{Calculation of the Action}

\noindent We calculate here the effective action $\Gamma[K]$ of the energy
history $\hat{K}(\br,t;\bv)={{1}\over{2}}v^2(\br,t)$. For each time $t$,
$\hK(\br,t)$ is a functional on phase space via its dependence on the random
initial data $\bv(\br)$ at $t=t_0$ of the Navier-Stokes solution $\bv(\br,t)$.
$K(t)$ is a possible value of this random variable, i.e. a numerical
time-history. According to the theorem established in \cite{Ey96I,Ey96II},
the effective action is characterized as the stationary point of the
``nonequilibrium action functional''
\be \Gamma[\hA,\hrho]=\int_{t_0}^\infty
dt\,\,\langle\hA(t),(\partial_t-\cL)\hrho(t)\rangle \lb{noneq-act} \ee
varied over arbitrary left and right ``trial functionals'' $\hA[\bv;t]$ and
$\hrho[\bv;t].$ (In \cite{Ey96II} these were
denoted $\Psi^L,\Psi^R$, respectively; the hat is used here to denote
functionals on the phase-space of velocity
fields). The variations are performed subject to the constraints of unit
overlap
\be \langle\hA(t),\hrho(t)\rangle=1 \lb{ovlap} \ee
and fixed mean (of $\hK(\br)$ {\em not} of $\hK(\br,t)$!)
\be \langle\hA(t),\hK(\br)\hrho(t)\rangle=K(t),\lb{fixmean} \ee
with the initial condition
\be \hrho[\bv;t_0]=\hat{P}_0[\bv], \lb{init-cond} \ee
where $\hat{P}_0$ is the initial Gaussian distribution at $t=t_0$, and with the
final condition
\be \hA[\bv;+\infty]\equiv 1. \lb{fin-cond} \ee
Note that $\langle\hA,\hrho\rangle=\int {\cal D}\bv \hA[\bv]\hrho[\bv]$. The
trial functional $\hrho(t)$ should be
taken to vary over the space of all probability distributions, while $\hA(t)$
is varied over the space of all bounded
observables. In this case the constraints become, more simply,
\be \langle \hA(t)\rangle_{\hrho(t)}=1 \lb{ovlap2} \ee
and
\be \langle \hA(t)\hK\rangle_{\hrho(t)} = K(t), \lb{fixmean2} \ee
in which $\langle\cdots\rangle_{\hrho(t)}$ denotes the average over the
distribution $\hrho(t)$. Likewise,
\be \Gamma[\hA,\hrho]= -\int_{t_0}^\infty dt\,\,\langle (\partial_t
+\cL^\dagger)\hA(t)\rangle_{\hrho(t)} \lb{noneq-act2} \ee
is a generally more useful expression for the nonequilibrium action.

To obtain the exact effective action, trial functionals should be varied over
the full spaces. However, within the Gaussian
{\em Ansatz} above, the variation is taken over a restricted class of trial
functionals. The right functional is just
the Gaussian PDF itself:
\be  \hrho[\bv;t]={{1}\over{\cN(t)}}\exp\left[-{{1}\over{2}}
                  \int d^3\bk
\,\,\hat{v}_i^*(\bk)\left(E^{-1}\right)_{ij}(\bk,t)\hat{v}_j(\bk)\right]
\lb{R-trial} \ee
with the isotropic spectral tensor
\be E_{ij}(\bk,t)= {{E(k,t)}\over{4\pi
k^2}}\left(\delta_{ij}-{{k_ik_j}\over{k^2}}\right) \lb{spec-tens} \ee
in which $E(k,t)$ is the scalar spectrum of Eq.(\ref{spectrum}) for a variable
total energy $E(t)$.  $\cN(t)$ is
the normalization factor guaranteeing total probability equal to unity. The
left trial functional within the Gaussian
PDF closure is chosen from among arbitrary linear combinations of the moment
functional $\hK[\br;\bv]$, which appeared
in the closure, and the constant functional $\equiv 1$:
\be \hA[\bv;t]=\alpha_0(t)1+\alpha_1(t)\hK[\br;\bv]. \lb{L-trial} \ee
The variable functions of time, $E(t),\alpha_0(t),\alpha_1(t)$ are the trial
parameters of the variational
calculation.

However, because of the two constraints, Eqs.(\ref{ovlap2}),(\ref{fixmean2}),
only one of these parameters
is independent. We shall take it to be $E(t)$. The unit overlap condition
Eq.(\ref{ovlap2}) requires that $\alpha_0(t)
+\alpha_1(t)E(t)=1$, or that
\be \hA[\bv;t]=1+\alpha_1(t)\left(\hK[\br;\bv]-E(t)\right) \lb{L-trial-II} \ee
by eliminating $\alpha_0(t)$. Next $\alpha_1(t)$ may be eliminated by using the
condition Eq.(\ref{fixmean2}). Since
\be \langle\hA(t)\hK(\br)\rangle_{E(t)}=
E(t)+\alpha_1(t)\left(\langle\hK^2(\br)\rangle_{E(t)}-E^2(t)\right),
    \lb{meanK} \ee
the constraint equation is obtained from an easily calculated average over the
Gaussian ensemble parametrized by $E(t)$.
Using $\langle v_i(\br)v_j(\br)\rangle_{E(t)}={{2}\over{3}}\delta_{ij}E(t)$,
this average is found to be
\be \langle\hK^2(\br)\rangle_{E(t)}={{5}\over{3}}E^2(t). \lb{sq-K} \ee
Note that evaluation  of this 1-point moment is the only place where Gaussian
statistics is employed in the whole
calculation. From the imposed condition Eq.(\ref{fixmean2}) we then obtain that
\be \alpha_1(t)={{3}\over{2}}E^{-2}(t)\left[K(t)-E(t)\right]. \lb{alpha1} \ee

The action may now be approximated as
\begin{eqnarray}
\Gamma_*[K;E] & = & -\int_{t_0}^\infty
dt\,\,\langle(\partial_t+\cL^\dagger)\hA(t)\rangle_{E(t)} \cr
          \,& = & \int_{t_0}^\infty
dt\,\,\left[-\dot{\alpha}_1(t)\langle\hK-E(t)\rangle_{E(t)}
+\alpha_1(t)\left(\dot{E}(t)
+\langle\hat{\varepsilon}\rangle_{E(t)}\right)\right] \cr
\,& = & \int_{t_0}^\infty dt\,\,\alpha_1(t)\left[\dot{E}(t)+\Lambda_m
E^p(t)\right] \cr
          \,& = & {{3}\over{2}}\int_{t_0}^\infty
dt\,\,E^{-2}(t)\left[K(t)-E(t)\right]
\left[\dot{E}(t)+\Lambda_m E^p(t)\right],
\lb{effact-EK}
\end{eqnarray}
in which $E(t)$ remains as the only trial parameter. We wrote as
$\hat{\varepsilon}(\br)={{\nu}\over{2}}\sum_{ij}
(\partial_iv_j(\br)+\partial_jv_i(\br))^2$ the local energy dissipation rate
and noted its average from Eq.(\ref{diss})
as $\langle\hat{\varepsilon}\rangle_{E(t)}=\Lambda_m E^p(t)$. By requiring
stationarity of the action under variations of
$E(t)$, or, $\delta\Gamma_*[K;E]/\delta E(t)=0$, with fixed $K(t)$, it is
straightforward to derive the variational equation
\be \Lambda_m E^p(t)+\dot{K}(t)=(p-2)\Lambda_m (K(t)-E(t))E^{p-1}(t).
\lb{var-parm} \ee
For any rational value of $p={{k}\over{l}}$, $k,l$ integers, this is a
polynomial of degree $k$ in $X=E^{1/l}$:
$(p-1)\Lambda_m X^k-(p-2)\Lambda_m K X^{k-l}+\dot{K}=0$. For a physically
allowable energy history, $K(t)>0$ and
$\dot{K}(t)<0$. Furthermore, $p>1$ whenever $m>-3$. Thus---independent of the
sign of $(p-2)$---the polynomial has one
change of sign in its coefficients for any permissable energy history.
Therefore, it follows from Descartes' rule of signs
that there is exactly one positive root $E(t)$ for each physical choice of
$K(t)$, when $p$ is rational. Because these
are dense in the real $p>0$, $E$ is uniquely defined for all permissable $K$.
Substituting that value into the
Eq.(\ref{effact-EK}) above, we obtain the final form of the approximate
effective action
\be \Gamma_*[K]={{3}\over{2(p-2)\Lambda_m}}\int_{t_0}^\infty dt \,\,
                    {{\left(\dot{K}(t)+\Lambda_m E^p(t)\right)
                    \left(\dot{E}(t)+\Lambda_m E^p(t)\right)}\over{E^{p+1}(t)}}
\lb{Gauss-eff} \ee
in which the $E$-dependence is eliminated by inserting the root of
Eq.(\ref{var-parm}) as described.

It is easy to check that, if the approximate action is evaluated at the
predicted closure mean energy $K_*(t)$, then
$\Gamma_*[K_*]=0$. In fact, using $\dot{K}_*(t)= -\Lambda_m K^p_*(t)$, the
corresponding $E_*(t)$ is determined from
\be
\Lambda_m\left(E_*^p(t)-K_*^p(t)\right)=
(p-2)\Lambda_m\left(K_*(t)-E_*(t)\right)E_*^{p-1}(t). \lb{Kstar-eq} \ee
The solution of this equation is
\be      E_*(t)=K_*(t)\,\,\,\,\&\,\,\,\,\dot{E}_*(t)=-\Lambda_m E_*^p(t).
\lb{Kstar} \ee
Obviously, substituting these values makes the approximate action vanish
identically. It can,
in fact, be shown that the mean value for {\em any} closure is a zero of the
approximate action evaluated by the
Rayleigh-Ritz method within that same closure \cite{Ey96I,Ey96II}. It may even
be shown further that the mean value
is always a stationary point of the action, $\delta\Gamma_*[K_*]/\delta
K(t)=0$. However, it need not be a minimum point,
as required by the realizability conditions on the effective action.

To examine the issue of realizability here, we consider small perturbations
$K(t)=K_*(t)+\delta K(t)$ from the predicted
mean. Because the calculation is straightforward but somewhat tedious, we give
the details in Appendix I. The final result
is that
\be \Gamma_*[K]={{3}\over{8(p-1)\Lambda_m}}\int_{t_0}^\infty dt
                          \,\,{{\left(\delta\dot{K}(t)+\Lambda_m
pK^{p-1}_*(t)\delta K(t)\right)^2}
                \over{K^{p+1}_*(t)}} + O\left((\delta K)^3\right).
\lb{quad-eff}
\ee
Note again that the coefficient $(p-1)$ in front of the action is $>0$ as long
as $m>-3$. In fact, $m>-1$ is required to
give a finite energy. Thus, for all permissable values of $m$, the approximate
action $\Gamma_*[K]$ satisfies
realizability, at least in a small neighborhood of the mean energy history
$K_*(t)$. One should be cautioned that
satisfaction of realizability is only a consistency check and cannot guarantee
correctness of predictions. Indeed, the
same calculation as we made above would carry through exactly for the 1D
Burgers equation, since the only property of
the nonlinear dynamics that was used was energy conservation. However, not all
of the previous results are true for
Burgers turbulence. In that case the energy spectrum Eq.(\ref{spectrum}) is not
even the correct quasi-equilibrium
form but, instead, a $k^{-2}$ spectrum will develop \cite{Burgers}. This
graphically illustrates that realizability
is perfectly compatible with falsity. On the other hand, we expect that the
approximation to the effective action with
the Kolmogorov spectrum Eq.(\ref{spectrum}) is qualitatively correct for 3D
Navier-Stokes turbulence. Unfortunately,
we have not so far been able to show that the full action,
Eq.(\ref{Gauss-eff}), satisfies all realizability constraints
for arbitrarily large deviations $\delta K$.

It may be observed from Eq.(\ref{quad-eff}) that the quadratic part of the
approximate action has precisely the form of
an Onsager-Machlup action \cite{OM}. Hence, it follows from the work of Onsager
and Machlup that the same law of
fluctuations is realized with the Langevin equation
\be \delta\dot{K}^+(t)+\Lambda_m pK^{p-1}_*(t)\delta K^+(t)=
(2R_*(t))^{1/2}\eta(t) \lb{Lang-eq} \ee
obtained by linearization of the energy-decay equation around its solution
$K_*(t)$ and by addition of a white-noise
random force $\eta(t)$, $\langle\eta(t)\eta(t')\rangle=\delta(t-t')$, with a
coefficient
\be R_*(t)={{2(p-1)}\over{3}}\varepsilon_*(t)K_*(t). \lb{FDT} \ee
This alternative stochastic representation is equivalent in the sense that all
finite distributions of $\delta K^+(t)$ in the
above Langevin model agree with those predicted for $\delta\hK(t)$ by the
quadratic action. \footnote{Recall from
\cite{Ey96I,Ey96II} that $\Gamma[K]$ is a generating functional for irreducible
multitime correlation functions of the
energy $\hK(t)$. See also Section 3.} We emphasize that this linear Langevin
representation is only adequate for the smaller
fluctuations about the mean and will not be sufficient to describe the larger
fluctuations. In fact, the quadratic part of
the
action is only a valid approximation for small deviations $\delta K$. The
predicted decay of the smaller energy fluctuations
according to a linearized law is similar to the Onsager regression hypothesis
for equilibrium fluctuations \cite{Ons31}.
Likewise, the expression Eq.(\ref{FDT}) is a {\em fluctuation-dissipation
relation} (FDR) analogous to that in equilibrium.
The white-noise term on the righthand side of Eq.(\ref{Lang-eq}) represents a
stochastic backscatter contribution to the
energy evolution and the Eq.(\ref{FDT}) relates its magnitude to the mean
energy dissipation rate $\varepsilon_*(t)$.
These are testable predictions of the closure hypotheses. We expect that the
prefactor $C={{2}\over{3}}(p-1)$ in the FDR,
whose precise value results from the Gaussian {\em Ansatz}, is correct at least
on order of magnitude. For $m=2$ its value
is $C={{5}\over{9}}\approx 0.56$.

The Gaussian {\em Ansatz} is obviously inadequate in one respect, because it
fails to capture the important non-Gaussian
effect of scale energy transfer. Let $\hat{\Pi}(k)$ be the usual spectral flux
as an {\em instantaneous} variable in
individual realizations, written in terms of a triple product of velocity
Fourier coefficients. (For example, see \cite{Ey94}
for an explicit expression). Then, one expects for freely-decaying turbulence
in the quasi-steady regime that $\langle
\hat{\Pi}(k)\rangle_{E(t)}=\varepsilon(t)$ for all inertial-range wavenumbers
$k_L(t)\ll k\ll k_d(t)$, breaking time-reversal symmetry. However, within
the Gaussian {\em Ansatz} $\langle\hat{\Pi}(k)\rangle_{E(t)}=0$. This pathology
of the Gaussian {\em Ansatz} shows up if
one calculates $\Gamma_*[\Pi]$, the Gaussian approximation to the effective
action of the flux $\hat{\Pi}(k)$.
In fact, $\Gamma_*[\Pi]$ is the Legendre transform of an approximate
cumulant-generating functional
$\lambda_*[H]$, using the notations of \cite{Ey96II}. That is,
$\Gamma_*[\Pi]=\max_H\left(H\Pi-
\lambda_*[H]\right)$. Because the 5th-order moment
$\langle\hK(\br)\hat{\Pi}(k)\rangle_{E(t)}=0$ in
the Gaussian {\em Ansatz}, as well as  $\langle\hat{\Pi}(k)\rangle_{E(t)}=0$,
it follows by the methods discussed
in \cite{Ey96II} that $\lambda_*[H]\equiv 0$ for all $H$. Therefore, the
Legendre transform is
\be \Gamma_*[\Pi]= \left\{\begin{array}{ll}
                                   0 & \Pi=0 \cr
                                   +\infty & \Pi\neq 0
                                  \end{array} \right. \lb{Gauss-VT} \ee
This result just implies that, within the Gaussian {\em Ansatz}, the flux
function $\hat{\Pi}(k)$ is identically
zero in every realization and no fluctuations from that value may occur. This
is clearly an unphysical result
of the closure. However, the failure of the Gaussian {\em Ansatz} to describe
the energy transfer is hoped not to
drastically affect the result for the effective action $\Gamma[K]$ of the
energy, because the model spectrum
Eq.(\ref{spectrum}) has built-in the correct overall decay rate.

Some insight into this may be obtained by considering the {\em exact} equation
for the 2-time correlation of
the energy fluctuation. The energy density fluctuation $\delta
\hK(\br,t)=\hat{K}(\br,t)-\langle\hat{K}(t)\rangle$
in each individual realization obeys the equation
\be \delta \dot{\hK}= -\grad\bdot[(\hat{K}+\hp)\bv-\nu\grad
\hat{K}]-\delta\hvepsilon. \lb{exact-eq} \ee
Here $\delta\hvepsilon =\hvepsilon-\langle\hvepsilon\rangle$ is the energy
dissipation fluctuation and $\hp$ is the
kinematic pressure. It is this equation which is being statistically modeled by
the Langevin equation, Eq.(\ref{Lang-eq}).
If this model is to be valid for second order statistics, then it must be true
that the exact equation
\be \langle \delta \dot{\hK}(t)\delta \hK(t_0)\rangle =
-\langle\grad\bdot[(\hat{K}+\hp)\bv
-\nu\grad \hat{K}](t)\delta \hK(t_0)\rangle-\langle\delta\hvepsilon(t)\delta
\hK(t_0)\rangle \lb{exact-corr} \ee
coincides with the one derived from the Langevin equation. For $t>t_0$ this is
just
\begin{eqnarray}
\langle\delta \dot{K}^+(t)\delta K^+(t_0)\rangle & = &
                \langle[-L_*(t)\delta K^+(t)+(2R_*(t))^{1/2}\eta(t)]\delta
K^+(t_0)\rangle \cr
\,  & = & -L_*(t)\langle\delta K^+(t)\delta K^+(t_0)\rangle. \lb{Lang-corr}
\end{eqnarray}
We have introduced $L_*(t)=\Lambda_m pK^{p-1}_*(t)$ and noted that the
white-noise force is uncorrelated
with earlier values of the energy fluctuation. In order to coincide with this
model equation, it is clear that
the first term due to space transport on the LHS of Eq.(\ref{exact-corr})
should be negligible, i.e. $\langle\grad\bdot
[(\hat{K}+\hp)\bv-\nu\grad \hat{K}](t)\delta \hK(t_0)\rangle\approx 0$. This is
plausible, because the space
transport term is rapidly varying in time and thus decorrelated with the energy
fluctuation at earlier time.
It is for this reason that such higher moments (of 4th and 5th order in
velocity) were never explicitly modeled
in our analysis, although they are implicitly represented by the white-noise
term in the Langevin equation. The
remaining term in the exact equation coincides with that in the model equation,
if it is further assumed that
\be  \langle\delta\hvepsilon(t)|\hat{K}(s),s< t\rangle \approx
-L_*(t)\delta\hK(t).  \lb{reg-hyp} \ee
That is, the conditional expectation of the dissipation fluctuation given the
value of the energy density over the
entire past should be obtained by {\em linearizing} the expression for mean
dissipation in the closure and evaluating
it at the present value of the energy fluctuation in the given realization.
This conditional relation is assumed to hold
when the energies $\hat{K}(t)$ are small deviations from the mean value
$\langle \hat{K}(t)\rangle\approx K_*(t)$.
This formula also has considerable plausibility: it is a formal statement of
the ``regression hypothesis'' on small
fluctuations. If this relation is used to eliminate $\delta\hvepsilon(t)$ in
Eq.(\ref{exact-corr}), then an equation of the
same form as Eq.(\ref{Lang-corr}) is obtained:
$\langle\delta\dot{\hK}(t)\delta\hK(t_0)\rangle\approx -L_*(t)\langle\delta
\hK(t)\delta \hK(t_0)\rangle.$ This should make more transparent the nature of
the approximations in the Rayleigh-Ritz
calculation at the level of closure considered here.

In fact, it is possible by such arguments to completely ``rederive'' the
Langevin model. If one accepts (i) the hypothesis
that the rapidly changing terms are correctly modeled by a white noise, i.e.
\be -\grad\bdot[(\hat{K}+\hp)\bv-\nu\grad \hat{K}](t)-[\delta\hvepsilon(t)
     -\langle\delta\hvepsilon(t)|\hat{K}(s),s< t\rangle] \approx
(2R_*(t))^{1/2}\eta(t), \lb{noise-hyp} \ee
and (ii) the ``regression hypothesis'' in Eq.(\ref{reg-hyp}), then the exact
equation Eq.(\ref{exact-eq}) reduces
to the Langevin model Eq.(\ref{Lang-eq}). A Kolmogorov-style dimensional
analysis would yield for the noise strength
$R_*(t)=C \varepsilon_*(t)K_*(t)$ with $C$ some universal constant, to be
determined. The value of this constant
$C={{2}\over{3}}(p-1)$ resulting from the variational calculation with the
Gaussian {\em Ansatz} will furthermore be
shown below to be the unique choice to recover the relation Eq.(\ref{level}).
Since this relation is exact when
single-point statistics of velocity {\em are} Gaussian---which is known to be a
very good approximation---the Langevin
model can be entirely motivated by intuitive considerations. The variational
calculation is a systematic analytical
procedure yielding the same Langevin model, but also only approximate. The two
derivations are therefore very
complementary.

Improvements of the Gaussian {\em Ansatz} are likely to lead to better results
for the effective actions. For example,
the ``synthetic turbulence'' models of \cite{JLSS} are random velocity fields
which contain the correct energy
transfer and also some of the intermittency effects of real turbulent
velocities. Using such statistical models
within our Rayleigh-Ritz scheme should lead not only to a qualitatively correct
result for $\Gamma[\Pi]$ but also
to quantitatively better results for $\Gamma[K]$. In such improved closures new
``test functionals'' in addition to
the quadratic moment-functional $\hK(\br;\bv)$ must be considered to determine
the time-dependence of the additional
free parameters in the statistical {\em Ansatz}. For example, the energy flux
variable (a triple moment-functional) would
be a natural variable to add to the closure. The choice of the ``test
functionals'' is an equally important element
of the closure as is the choice of the model velocity statistics. We emphasize
again that our results above for $\Gamma[K]$
depend very little on the choice of the Gaussian statistics. For {\em any}
model statistics with the mean energy $E(t)$
as the only free parameter and with $\hK(\br;\bv)$ the corresponding ``test
functional'', results very similar to
those above will follow. In that general setting a result $\langle
\hK^2(\br)\rangle_{E(t)}=B E^2(t)$ will hold by
dimensional analysis, for some constant $B$, replacing Eq.(\ref{sq-K}). This
means that the formula Eq.(\ref{Gauss-eff})
for the approximate action will still hold, with the factor ${{3}\over{2}}$ in
front simply replaced by another
number $D=1/(B-1)$ of order unity. Only the value $B={{5}\over{3}}$ depends
upon the Gaussian {\em Ansatz}. By employing
improved closures one may hope to derive from first principles such theoretical
features as the ``permanence of large
eddies'' for $m<4$. Because the Rayleigh-Ritz algorithm is a convergent
approximation scheme for the true effective
actions, systematic improvement of the closures will lead to a refined
description of the turbulent dynamics.

\section{Testing the Theory}

\noindent The previous theory has testable consequences for turbulent energy
fluctuations. The most likely experimental
situation for such checks is grid turbulence, which well approximates a
homogeneous, isotropic, decaying turbulence. In
principle, it would be possible to make an experiment by measuring the velocity
at a single point $\br$ in the frame of
mean downstream motion. Constructing from this the energy history $\hK(t)\equiv
{{1}\over{2}}v^2(\br,t)$ and compiling
an ensemble of realizations, one may, as in \cite{C-BC,C-BCII}, compute various
multi-time statistics to compare
with the theory.

The most familiar such statistics are the $r$-time correlation functions
$\langle \hK(t_1)\cdots \hK(t_r)\rangle$.
These may be derived directly from the effective action $\Gamma[K]$. In fact,
by taking $r$ functional
derivatives of the action, evaluated at the mean value, the {\em irreducible}
$r$-time correlators are obtained:
\be \langle \hK(t_1)\cdots \hK(t_r)\rangle^{{\rm irr}}
                      = \left. {{\delta^r \Gamma[K]}\over{\delta
K(t_1)\cdots\delta K(t_r)}}\right|_{K=K_*}. \lb{irred} \ee
For this result, for the definition of irreducible correlators and their
relation to the connected correlators
(or cumulants), see any text in quantum field theory, e.g. \cite{Itz-Zub},
Section 6.2.2 or \cite{Huang}, Section 10.2.
We only note here that the irreducible 2-time correlator, or $\langle
\hK(t_1)\hK(t_2)\rangle^{{\rm irr}}$, is the inverse
operator kernel of the connected 2-time function (covariance) $\langle
\hK(t_1)\hK(t_2)\rangle^{{\rm con}}=\langle\delta
\hK(t_1)\delta \hK(t_2)\rangle$, i.e.
\be \int ds\,\,\langle \hK(t)\hK(s)\rangle^{{\rm irr}}\langle\delta
\hK(s)\delta \hK(t')\rangle=\delta(t-t').
\lb{inverse} \ee
Relations between higher-order irreducible and connected correlators are
obtained by taking further functional derivatives
of this relation with respect to $K$. See \cite{Itz-Zub,Huang}.

It is very easy to obtain the variational approximation $\langle
\hK(t)\hK(s)\rangle^{{\rm irr}}_*$ from Eq.(\ref{irred})
and the quadratic part of the Gaussian effective action, Eq.(\ref{quad-eff}).
Taking its inverse, the covariance
$\langle\delta \hK(t)\delta \hK(t')\rangle_*$ is then evaluated as
\begin{eqnarray}
\langle\delta \hK(t)\delta \hK(t')\rangle_* & = & \exp\left[-\int_{t_0}^t
ds\,\,L_*(s)-\int_{t_0}^{t'} ds\,\,L_*(s)\right](\delta K_0)^2  \cr
\, & &\,\,\,\,\,\,+ 2\int_{t_0}^{\min\{t,t'\}} ds\,\,R_*(s)\exp\left[-\int_s^t
dr\,\,L_*(r)-\int_s^{t'}dr\,\,L_*(r)\right]  \lb{predict}
\end{eqnarray}
in our theory. We have again written $L_*(t)=\Lambda_m pK^{p-1}_*(t)$. These
calculations are outlined
in Appendix II. In the same way, by taking an arbitrary number $r$ of
functional derivatives in Eq.(\ref{irred}), all
correlations of any finite order are obtainable from the effective action.
However, we do not pursue the
general calculation here.

To cast the theoretical results into a form that may be compared with
experiment, we insert the mean decay law $K_*(t)$
from Eq.(\ref{law}) into Eq.(\ref{predict}) and perform the integrals. For the
2-time covariance of the turbulent energy
this calculation is straightforward and the prediction is:
\begin{eqnarray}
\, & & \langle \delta \hK(t)\delta \hK(t')\rangle_*=
       \left({{t-t_0^*}\over{\Delta
t}}\right)^{-(n+1)}\left({{t'-t_0^*}\over{\Delta t}}\right)^{-(n+1)}\times \cr
\, & & \,\,\,\,\,\,\,\,\,\,\,\,\,\,\,\,\,\,\,\,\,\,\,\,\,\,
\,\,\,\,\,\,\,\,\,\,\,\,\,\,\,\,\,\,\,\,\,\,
\left\{ (\delta K_0)^2 +{{2}\over{3}} K_0^2
       \left[\left({{t_{{\rm min}}-t_0^*}\over{\Delta t}}\right)^2 -
1\right]\right\}, \lb{prediction}
\end{eqnarray}
with $t_{{\rm min}}=\min\{t,t'\}$. The notations are the same as for the mean
decay law. It should be noted
that the first term $\propto (\delta K_0)^2$ corresponds to decay of an initial
energy fluctuation $\delta K_0$.
The second term $\propto K_0^2$ represents the new fluctuations generated by
the internal turbulence noise, through the
stochastic backscatter dynamics. As a consequence of that term, the long-time
rms value of the energy, $K_{{\rm rms}}(t)
=\left[\langle(\delta \hK(t))^2\rangle_*\right]^{1/2}$, evolves to a constant
level with respect to the mean energy $K_*(t)$:
\be \lim_{t\rightarrow\infty}{{\langle(\delta
\hK(t))^2\rangle_*}\over{K_*^2(t)}}= {{2}\over{3}}. \lb{level} \ee
The limiting value of ${{2}\over{3}}$ is what would occur for an asymptotic
Gaussian statistics of the 1-point velocity
variable. If we had adopted a non-Gaussian {\em Ansatz} in our calculation,
then the predicted limiting value would have
been $1/D=(B-1)$. Since any value of the constant can be accommodated by an
appropriate such {\em Ansatz}, it is not so
important to the theory which particular constant is correct (although the
Gaussian value is expected to be quite accurate).
What would falsify the present theory would be an experimental finding that the
functional form in Eq.(\ref{prediction})
was wrong, for any possible choice of the constant $1/D$ replacing
${{2}\over{3}}$. We should emphasize, however, that the
standard theory for the mean energy decay law $K_*(t)$ in Eq.(\ref{law}) is in
agreement with present experiments, with a
value of $n$ near $1.2$. In all cases studied so far, PDF {\em Ans\"{a}tze}
adequate to predict the mean values of selected
variables have also, employed in our Rayleigh-Ritz method, yielded good
predictions for the fluctuations of those variables
near the means. See \cite{EyAl}. We therefore expect the prediction in
Eq.(\ref{prediction}) to be reasonably accurate.

The previous result for $r=2$, the covariance in Eq.(\ref{predict}), may be
obtained as well from the Langevin
equation, Eq.(\ref{Lang-eq}). However, it must be emphasized that only the {\em
small} fluctuations of the energy, with
$\delta \hK(t)\ll K_*(t)$, are expected to be distributed according to that
linearized equation. Because correlation
functions will get sizable contributions from the larger fluctuations, for
which the linear law breaks down, it would not
be appropriate to compare general $r$-time correlation functions of $\hK(t)$
obtained from the linear theory with
experiment. It only happens for $r=2$ that the linear Langevin equation and the
full (nonlinear) effective action yield the
same predictions.

It is possible to give both the effective action functional and the linear
Langevin equation a direct
empirical significance in grid turbulence. This is based upon the standard
device of making $N$ independent trials
to calculate the averages from experiment. Indeed, performing the same decay
experiment $N$ times independently, one
usually considers an {\em empirical mean history}
\be \overline{K}_N(t)={{1}\over{N}}\sum_{i=1}^N \hK_i(t) \lb{mean-hist} \ee
formed from the realizations $\hK_i(t),\,\,\,i=1,2,...,N,$ of the $N$ different
samples. \footnote{Assuming that the
turbulence is indeed statistically homogeneous, these $N$ measurements might
even be taken from points $\br_i,\,\,i=1,...,N$
in the same flow but at separations greater than $L$, the integral scale, to
assure statistical independence.} The
effective action $\Gamma[K]$ measures the probability for the empirical mean
$\overline{K}_N(t)$ (which is a random quantity
at finite $N$) to take on a value very different from the true ensemble-average
$\langle\hK(t)\rangle$. More precisely,
\be {\rm Prob}\left(\{\overline{K}_N(t)\approx K(t): t_0\leq t<+\infty\}\right)
               \sim \exp\left( -N \Gamma[K]\right). \lb{LD} \ee
Thus, the probability to observe $\overline{K}_N(t)$ taking any value $K(t)$
other than the true ensemble-average
$\langle\hK(t)\rangle$ is exponentially small in the number of samples $N$.
This is a consequence of the famous Cram\'{e}r
theorem on large-deviations of sums of independent random variables (e.g. see
\cite{Frisch}, Section 8.6.4
and references therein). Put another way, the dimensionless quantity
$1/\Gamma[K]$ gives an estimate of the
number $N$ of additional independent samples required to reduce by $e$-fold the
probability of the fluctuation value $K$
in the empirical average $\overline{K}_N$. In principle, therefore, the
effective action $\Gamma[K]$ is itself directly
measurable in grid turbulence, by assembling a histogram of observed histories
$\overline{K}_N(t)$ and determining the
decay rate of the probabilities for large $N$. However, this would not be
feasible with a reasonable number of independent
samples $N$ except for histories $K(t)$ sufficiently near the mean history.

The linear Langevin model Eq.(\ref{Lang-eq}) is more restricted in its
validity, since, as has been stressed, it is
equivalent to the quadratic action and is adequate only to predict 2nd-order
statistics of $\hK(t)$. In particular,
a linear Langevin equation can produce only Gaussian multi-time statistics for
$\delta K^+(t)$. (Of course,
this has nothing to do with the use of a Gaussian PDF {\em Ansatz} for the {\em
velocity field} and will be true even
if a non-Gaussian {\em Ansatz} is employed: see Appendix I.) The true
statistics of $\delta \hK(t)$ will not be Gaussian
at all, e.g. they will be ``chi-square'' if the 1-point velocity itself is
Gaussian. Nevertheless, the Langevin equation
can also be given a direct empirical significance in terms of the independent
$N$-sample ensemble, based upon the central
limit theorem. It accurately describes the statistics of the normalized sum
variable:
\be \delta{\hK}_N(t)={{1}\over{N^{1/2}}}\sum_{i=1}^N
(\hK_i(t)-\langle\hK(t)\rangle). \lb{clt-sum} \ee
This quantity has zero mean and the same covariance as $\delta
\hK_i(t)=\hK_i(t)-\langle\hK_i(t)\rangle$ for each
independent sample $i$, i.e.
$\langle\delta{\hK}_N(t)\delta{\hK}_N(t')\rangle
=\langle\delta{\hK}_i(t)\delta{\hK}_i(t')
\rangle$ for all $i$. However, it is furthermore a Gaussian variable in the
limit of large $N$, in agreement with the
linear Langevin equation. Thus, at a large but finite $N$ it is legitimate to
compare predictions of the correlations of
$\delta K^+(t)$ using the linear Langevin dynamics, Eq.(\ref{Lang-eq}), with
those from experiment for $\delta \hK_N(t)$
at large $N$. Notice that the Gaussian statistics for $\delta\hK_N(t)$ in fact
result by substituting into $\Gamma[K]$
in Eq.(\ref{LD}) the value $K=\langle\hK\rangle+N^{-1/2}\delta K$. In that
case, by expanding in $\delta K$, one obtains
\be {\rm Prob}\left(\{\delta{\hK}_N(t)\approx \delta K(t):t_0\leq
t<+\infty\}\right)
               \sim \exp\left( -\Gamma_{(2)}[\delta K]\right), \lb{LD-quad} \ee
where $\Gamma_{(2)}[\delta K]$ is the quadratic approximation to the exact
effective action and terms in the exponent
of order $N^{-1/2}$ have been neglected in the large $N$ limit. This is just
one of the standard proofs of the central
limit theorem. The important point here is that it naturally accounts why the
linear Langevin equation may be appropriate
to calculate 2nd-order statistics but certainly not higher order. For the
latter purpose the full nonlinear action,
Eq.(\ref{Gauss-eff}), must be used, not just the quadratic part.

\section{Conclusions}

\noindent The main results of this work are as follows:

(1) We have derived an action functional, Eq.(\ref{Gauss-eff}), for energy
histories in decaying homogeneous
and isotropic turbulence at high Reynolds number, by means of a Rayleigh-Ritz
calculation using standard closure
assumptions. This action generalizes the Onsager-Machlup action to
fully-developed turbulent flow and characterizes
the mean energy history by a variational principle.

(2) We have shown that the quadratic part of the action, Eq.(\ref{quad-eff}),
valid for a region of small
fluctuations sufficiently near the mean, satisfies all required realizability
constraints. In fact, it is of
the Onsager-Machlup form and thus has a stochastic realization by a linear
Langevin equation for the energy history.
The deterministic part of the equation is obtained by linearization of the mean
decay law and the random part has
its strength determined by a ``fluctuation-dissipation relation'' in terms of
the mean energy dissipation.

(3) Testable consequence of the theory are $r$-time correlation functions of
the energy, which may be obtained
by functional differentiation of the effective action. These are not obtainable
from the starting hypotheses
by direct PDF methods. As an example, the $2$-time cumulant, or covariance, of
the energy history is derived
in detail.

(4) A direct empirical significance of the effective action is given in terms
of fluctuation probabilities
for ensemble averages over $N$ independent samples or ensemble-points. This
interpretation permits the
effective action itself to be measured experimentally, at least for arguments
in path-space sufficiently close
to the mean history.

\vspace{0.3in}

\noindent {\bf Acknowledgements.} I wish to thank U. Frisch, N. Goldenfeld, Y.
Oono and D. Stein for conversations
some years ago which helped me to realize that nonequilibrium action principles
should have important applications
to strongly noisy systems such as turbulence. B. Bayly also asked a very
stimulating question regarding
the operational interpretation of the effective action and R. H. Kraichnan has
greatly encouraged the work
with his interest and ideas.

\newpage

\section {Appendices}

\noindent {\bf Appendix I: Quadratic Approximation to the Action}

\noindent Let us consider in Eq.(\ref{Gauss-eff}) small fluctuations $\delta
K,\delta E$:
\be  K(t)=K_*(t)+\delta K(t)\,\,\,\,\&\,\,\,\,E(t)=K_*(t)+\delta E(t),
\lb{fluc} \ee
using the fact that $E_*(t)=K_*(t)$. It is then easy to see from
Eq.(\ref{Gauss-eff}) that, up to quadratic order,
\be \Gamma_*[K]={{3}\over{2(p-2)\Lambda_m}}\int_{t_0}^\infty dt
                          \,\,{{\left(\delta\dot{K}(t)+\Lambda_m
pK^{p-1}_*(t)\delta E(t)\right)
                               \left(\delta\dot{E}(t)+\Lambda_m
pK^{p-1}_*(t)\delta E(t)\right)}
                \over{K^{p+1}_*(t)}}. \lb{quad-eff2}
\ee
By a straightforward linearization of the determining Eq.(\ref{var-parm}), it
follows that
\be -2(p-1)\Lambda_m K_*^{p-1}(\delta E-\delta
K)=\delta\dot{K}+\Lambda_mpK_*^{p-1}\delta K. \lb{linear} \ee
Let us introduce a shorthand notation for the lefthand side of this equation:
\be \Delta\equiv \delta\dot{K}+\Lambda_mpK_*^{p-1}\delta K. \lb{Delta} \ee
Using the Eq.(\ref{linear}), it is not hard to show that
\be \delta\dot{K}(t)+\Lambda_m pK^{p-1}_*(t)\delta
E(t)={{(p-2)}\over{2(p-1)}}\Delta \lb{first} \ee
and
\be \delta\dot{E}(t)+\Lambda_m pK^{p-1}_*(t)\delta E(t)={{1}\over{2}}\Delta
-K_*{{d}\over{dt}}\left({{\Delta}\over{2(p-1)\Lambda_m K_*^p}}\right).
\lb{second} \ee
Substituting these into Eq.(\ref{quad-eff2}) above, it follows that, to
quadratic order,
\begin{eqnarray}
\Gamma_*[K] & = & {{3}\over{2(p-2)\Lambda_m}}\int_{t_0}^\infty dt
                          \,\,{{(p-2)}\over{2(p-1)}}{{\Delta}\over{K_*^{p+1}}}
\left[{{1}\over{2}}\Delta-K_*{{d}\over{dt}}\left({{\Delta}\over{2(p-1)\Lambda_m
K_*^p}}\right)\right] \cr
                 \,& = & {{3}\over{8(p-1)\Lambda_m}}\int_{t_0}^\infty
dt\,\,{{\Delta^2(t)}\over{K_*^{p+1}(t)}}
                         -{{3}\over{16(p-1)^2\Lambda_m^2}}\int_{t_0}^\infty
dt\,\,{{d}\over{dt}}\left[\left({{\Delta}\over{K_*^p}}\right)^2\right] \cr
                \,& = & {{3}\over{8(p-1)\Lambda_m}}\int_{t_0}^\infty
dt\,\,{{\Delta^2(t)}\over{K_*^{p+1}(t)}}.
\lb{final}
\end{eqnarray}
To obtain the last line we used the boundary conditions
$\Delta({t_0})=\Delta(\infty)=0$, which are required by
the Eqs.(\ref{init-cond}), (\ref{fin-cond}), (\ref{alpha1}). It should be
obvious that the last line of
Eq.(\ref{final}) is the same as the result, Eq.(\ref{quad-eff}), claimed in the
text.

Although the present calculation employed the Gaussian {\em Ansatz}, it should
be stressed that a similar result
will hold for more realistic statistical models of the velocity field. In
particular, the quadratic form of the
action does not depend upon the Gaussian assumption, but is simply a
consequence of the fact that the
mean history $K_*(t)$ is required to be an absolute minimum. Hence, this will
be true for any closure model
leading to an effection action satisfying the realizability conditions. In that
case, an expansion in small
deviations $\delta K(t)$ around the mean must necessarily lead to a quadratic
expression involving the
linearized evolution expression,
$\Delta(t)=\delta\dot{K}+\Lambda_mpK_*^{p-1}\delta K$. What will be different
for other closures is the coefficient multiplying $\Delta^2(t)$, which
correspond to different predictions
of the fluctuations around the mean.

\newpage

\noindent {\bf Appendix II: The Predicted 2-Time Cumulant}

\noindent As observed in the text, the irreducible $r$-time correlations of
$\hK(t)$ can be obtained from
functional derivatives of $\Gamma[K]$ at $K=K_*$: see Eq.(\ref{irred}).
Equivalently, these irreducible
correlators can be read off from the Taylor series:
\be \Gamma[K]=\sum_{r=2}^\infty {{1}\over{r!}}\int dt_1\cdots\int dt_r
                                      \Gamma_r(t_1,\cdots,t_r)\delta
K(t_1)\cdots \delta K(t_r), \lb{Taylor} \ee
as $\langle \hK(t_1)\cdots \hK(t_r)\rangle^{{\rm
irr}}=\Gamma_r(t_1,\cdots,t_r)$. It is thus easy to obtain
in the Gaussian approximation $\langle \hK(t_1)\hK(t_2)\rangle^{{\rm irr}}_*$
from the quadratic part
of the Gaussian action derived in the previous
Appendix I, written as
\be \Gamma_*[K]={{1}\over{2}}\int dt_1\int dt_2
                          \,\,{{\left(\delta\dot{K}(t_1)+L_*(t_1)\delta
K(t_1)\right)
                               \left(\delta\dot{K}(t_2)+L_*(t_2)\delta
K(t_2)\right)}\over{2R_*(t_1)}} \delta(t_1-t_2).
\lb{quad-eff3}
\ee
Thence,
\be \langle \hK(t_1)\hK(t_2)\rangle^{{\rm irr}}_*
=[-\partial_{t_1}+L_*(t_1)]\left(2R_*(t_1)\right)^{-1}[\partial_{t_1}+L_*(t_1)]
                                            \delta(t_1-t_2) \lb{irred-2} \ee
and, from Eq.(\ref{inverse}), the 2-time cumulant must satisfy
\be
[-\partial_{t_1}+L_*(t_1)]\left(2R_*(t_1)\right)^{-1}[\partial_{t_1}+L_*(t_1)]
         \langle\delta \hK(t_1)\delta \hK(t_2)\rangle_* =\delta(t_1-t_2).
\lb{cum-eq} \ee
To solve this equation, we use the Greens functions
\be [-\partial+L_*]^{-1}(t,t')=\exp\left[\int_{t'}^t
ds\,\,L_*(s)\right]\theta(t'-t) \lb{Green1} \ee
\be [\partial+L_*]^{-1}(t,t')=\exp\left[-\int_{t'}^t
ds\,\,L_*(s)\right]\theta(t-t'). \lb{Green2} \ee
which are anti-causal and causal, resp. Performing one integration we obtain,
for $s\leq t_2$,
\begin{eqnarray}
[\partial_s+L_*(s)]\langle\delta \hK(s)\delta \hK(t_2)\rangle_*
                            & = & 2R_*(s)[-\partial_s+L_*(s)]^{-1}\delta(s-t_2)
\cr
                          \,& = & 2R_*(s)\exp\left[-\int_s^{t_2}
dr\,\,L_*(r)\right]\theta(t_2-s) \cr
                          \,& \equiv & G(s;t_2). \lb{1st-int}
\end{eqnarray}
A second integration forward from ${t_0}$ gives, for $t_1\leq t_2$,
\be \langle\delta \hK(t_1)\delta \hK(t_2)\rangle_*=\exp\left[-\int_{t_0}^{t_1}
ds\,\,L_*(s)\right]F(t_2)
                     +\int_{t_0}^{t_1} ds\,\,\exp\left[-\int_s^{t_1}
dr\,\,L_*(r)\right]G(s;t_2). \lb{2nd-int} \ee
The first term on the right is an arbitrary solution of the homogeneous
equation. The function $F(t_2)$ is
determined by symmetry and initial conditions as
\be F(t_2)=\exp\left[-\int_{t_0}^{t_2} ds\,\,L_*(s)\right]\langle \left(\delta
\hK(t_0)\right)^2\rangle. \lb{ic} \ee
When this is substituted into Eq.(\ref{2nd-int}) along with the definition of
$G(s;t_2)$ from Eq.(\ref{1st-int}),
the claimed result, Eq.(\ref{predict}) in the text, is obtained. Note that
$\langle \left(\delta \hK(t_0)\right)^2\rangle$
defines the quantity $(\delta K_0)^2$ in Eq.(\ref{predict}) of the text.

It is easy to check that this result coincides also with $\langle\delta
K^+(t)\delta K^+(t_0)\rangle$ calculated from
the Langevin model, Eq.(\ref{Lang-eq}). However, in general, for correlations
of order $r>2$, the Langevin model is not
adequate and the above method of calculation from the effective action must be
employed.

\end{document}